\documentclass[epj]{svjour}

\usepackage{graphicx}
\usepackage{fancyhdr}
\usepackage[centertags]{amsmath}
\usepackage{amssymb}
\usepackage{epsfig}
\usepackage{subfigure}
\usepackage{color}
\usepackage{pifont}
\usepackage{enumerate}

\def\makeheadbox{{
 }}
\def\mytitle{My title} 
\def\myauthors{My name}  
\def\mytype{My type of session}
\def\mysession{My session}

\def\mytitle{Do Particles in Incomplete GUT Multiplets Always Spoil Unification?} 
\def\myauthors{Ak{\i}n Wingerter}    
\def\mytype{Contributed Talk}    
\def\mysession{Theoretical Models}

\newcommand{\Red}[1]{{\color{red}{#1}}}
\newcommand{\bs}[1]{\boldsymbol{#1}}
\newcommand{\bsb}[1]{\boldsymbol{\overline{#1}}}
\newcommand{\ra}{\rightarrow}

\newcommand{\SU}[1]{\text{SU}(#1)}

\newcommand{\U}[1]{\text{U}(#1)}
\newcommand{\Hc}{\bar{H}}

\newcommand{\uc}{\bar{u}}
\newcommand{\dc}{\bar{d}}
\newcommand{\ec}{\bar{e}}
\newcommand{\nc}{\bar{\nu}}
\newcommand{\m}{\text{-}}
\newcommand{\q}{\phantom{\text{-}}}
\newcommand{\gut}{{\textsc{gut}}}
\newcommand{\susy}{{\textsc{susy}}}
\newcommand{\alphagut}{\alpha_{\gut}}
\newcommand{\alphaem}{\alpha_{\text{em}}}

\newcommand{\mgut}{{M_{\gut}}}

\begin{document}
\title{Do Incomplete GUT Multiplets Always Spoil Unification?}
\subtitle{}
\author{Ak{\i}n Wingerter
}                     
%
%
\institute{The Ohio State University, Physics Department, 191 W.~Woodruff Ave, Columbus, OH 43202}
%
\date{}
\abstract{
We consider a new class of light vectorlike exotics with fractional electric charge which do {\it not come in complete representations} of a grand unified gauge group, and are nevertheless {\it compatible with gauge coupling unification} and other predictions from Grand Unified Theories. Such states naturally arise in orbifold constructions of the heterotic string. Some aspects of their phenomenology and the consequences for the LHC are explored.\\[1ex]
Submitted for the SUSY07 proceedings. Based on Phys.~Rev.~Lett.~{\bf 99}, 051802 (2007) in collaboration with Stuart Raby.
} 
\maketitle
%

\section{Motivation}
\label{intro}

\subsection{The Standard Model and Beyond}

There is no convincing experimental data in disagreement with the Standard Model (SM), and yet we have good reasons to believe in new physics beyond the scale of a few TeV: The Standard Model has 26-28 parameters (including Dirac or Majorana neutrinos) which seemingly take arbitrary values. There is no explanation for the observed pattern of gauge symmetries and no organizing principle for the particles in each generation, and also no reason why these particles come in 3 generations at all. The mass of the only scalar particle in the theory, the yet-to-be-discovered Higgs boson, receives large radiative corrections and requires an incredible fine-tuning to be of order the weak scale. Over the last decade we learned from cosmology that Standard Model particles can account only for $\sim$ 4\% of the matter and energy content of the universe, leaving it open for speculation what may constitute $\sim$ 23\% which is known to be non-baryonic matter.

\begin{table}
\renewcommand{\arraystretch}{1.4}
\caption{Spectrum of the (Minimal Supersymmetric) Standard Model. All states other than these are termed ``exotic''.}
\label{tab:smparticles}
\begin{center}
\begin{tabular}{|c|l||c|l||c|l|}
\hline
$Q$ & $(\bs{3}, \bs{2})_{\q 1/3}$       &   $L$ & $(\bs{1}, \bs{2})_{\m 1}$   &  $H$ & $(\bs{1}, \bs{2})_{\q 1}$     \\
$\uc$ & $(\bsb{3}, \bs{1})_{\m 4/3}$ &   $\ec$ & $(\bs{1}, \bs{1})_{\q 2}$    &  $\Hc$ & $(\bs{1}, \bs{2})_{\m 1}$\\
\cline{5-6}
$\dc$ & $(\bsb{3}, \bs{1})_{\q 2/3}$    &   $\nc$ & $(\bs{1}, \bs{1})_{\q 0}$    &  \multicolumn{2}{|c|}{}            \\
\hline
\end{tabular}
\end{center}
\end{table}

\subsection{Grand Unification and Supersymmetry}
\label{sec:gut_and_susy}

\begin{figure}
\begin{center}
    \subfigure{\epsfig{figure=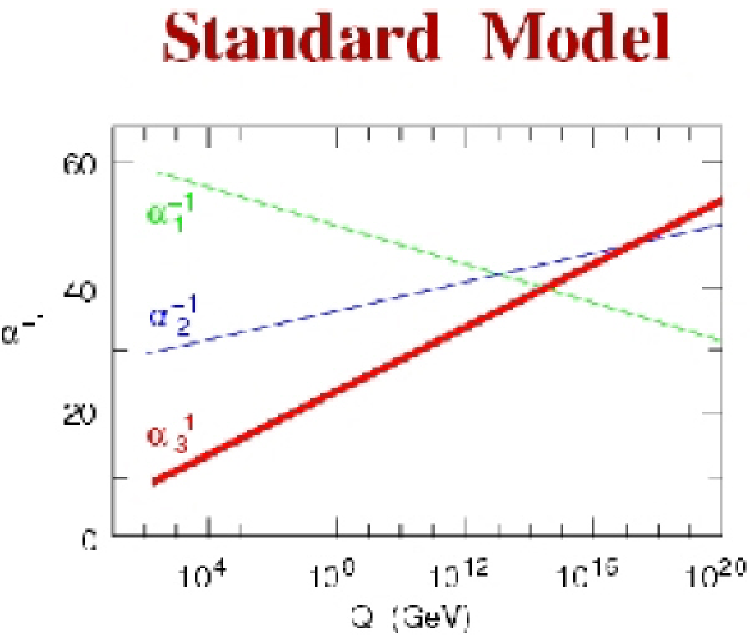, angle=0, scale=0.5}}
    \subfigure{\epsfig{figure=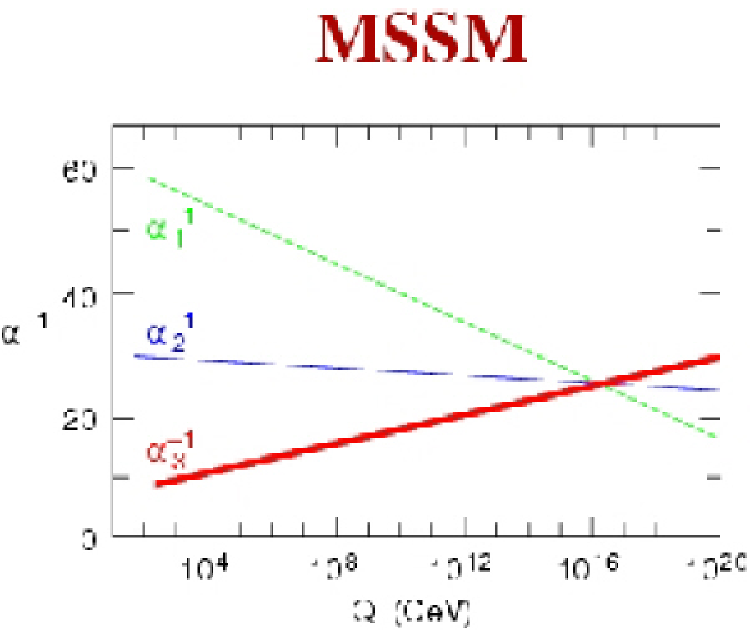, angle=0, scale=0.5}}
    \caption{The renormalization group running of the gauge couplings in the case of the SM and MSSM, respectively.}
    \label{fig:unif}
  \end{center}
\end{figure}

Are there hints at physics beyond the Standard Model? When we extrapolate the couplings measured at the weak scale $\sim 200$ GeV to higher energies, they {\it seem to meet} at one point. It turns out that in the case of the SM, the couplings miss each other, whereas in the case of the Minimal Supersymmetric Standard Model (MSSM), they {\it do unify} within the experimental and theoretical uncertainties (see Fig.\ref{fig:unif}). This is a strong indication not only for the existence of a Grand Unified Theory (\gut{}) at the scale of $\sim 3\times 10^{16}$ GeV ($\mgut$), but also for supersymmetry (\susy{}) with superpartner masses $\sim 1$ TeV \cite{drw,raby}.

\subsection{Exotic Particles}
\label{sec:exotic_particles}

By definition, all particles which are not in the MSSM spectrum (see Tab.~\ref{tab:smparticles}) are termed {\it exotics}. Most theories beyond the Standard Model predict the existence of exotic particles. Since no exotics have been observed so far, they must necessarily be heavier than the electroweak scale\footnote{The exact mass bounds depend on the quantum numbers of the exotic particles, their production mechanism, and other factors. See Ref.~\cite{Yao:2006px} current values.}. 

The running of the gauge coupling constants sensitively depends on the spectrum of the theory. Any exotic particles between the electroweak and $\mgut$ will {\it generally} spoil this nice picture of unification, {\it unless} these new particles come in complete representations of the unified gauge group. The simplest example would be an $\SU{5}$ as the group at the unification scale and a pair of $\bs{5}+\bsb{5}$ as exotic particles, which decompose into the SM representations
\begin{equation}
\renewcommand{\arraystretch}{1.4}
\begin{array}{ccccccccccc}
\bs{5} & + & \bsb{5} & \ra & (\bs{3},\bs{1})_{\m 2/3} & + & (\bs{1},\bs{2})_{\q 1} & + & (\bs{3},\bs{1})_{\q 2/3} & + & (\bs{1},\bs{2})_{\m 1}.\\
       &   &         &     & \dc^c & + & L^c & + & \dc & + & L
\end{array}
\label{eq:5_and_5bar}
\end{equation} 
In this case, the exotic particles are a fourth-generation $\dc$, $L$, and their charge conjugates $\dc^c$, $L^c$. (All fields are left-chiral. See Tab.~\ref{tab:smparticles} for the nomenclature.) If e.g.~$\dc$ and $\dc^c$ were the only ``extra'' particles in the theory, the gauge couplings would not unify.

\subsection{Vectorlike Pairs}

Gauge symmetries do not allow for explicit mass terms for chiral particles in the Lagrangian, and the well-known Higgs mechanism generates an effective mass of order the electroweak scale. For particles to naturally acquire a mass well above the electroweak scale, they must come in {\it vectorlike pairs}, i.e.~each state $\chi$ must be accompanied by its charge conjugate $\chi^c$. The mass term
\begin{equation}
M \chi \chi^c
\end{equation}
is then gauge invariant and can be much larger than 200 GeV. In general, there are 2 arguments which seem to indicate that this mass is rather of order the \gut{} scale. First, $\mgut$ is the only scale in the theory, so it is ``natural'' to expect $M$ to be of the same order. Second, any particles below $\mgut$ contribute to the running of the gauge couplings and may spoil unification. 

\medskip

The first argument is aesthetic in nature and more of a guideline for model builders than a real constraint. The second one confronts theory with data: If we believe in unification, there must not be any exotic particles lighter than $\mgut$, which affect the \gut{} relations.

It is widely assumed that the only exotic particles which can be lighter than $\mgut$ and do not change the predictions from Grand Unification are those in complete representations of the \gut{} group. 

In this study (see also Ref.~\cite{Raby:2007hm}), we argue that there is a more general class of exotics which are {\it not in complete representations} of the \gut{} group and nevertheless {\it do not change most of the predictions} from Grand Unification. In fact, their only effect is to increase the value of the gauge coupling $\alphagut{}$ at the \gut{} scale.

\section{A New Class of Exotic Particles}
\label{sec:new_class_of_exotics}

\subsection{The Renormalization Group Equations}

The renormalization group (RG) running of the gauge couplings at one loop is given by
\begin{equation}
\frac{1}{\alpha_i(\mu)} = \frac{1}{\alphagut} - \frac{1}{2\pi}\,b_i \log\left(\frac{\mgut}{\mu} \right),
\label{eq:rge}
\end{equation}
where the $i=1,2,3$ refers to the gauge groups $\U{1}_Y$, $\SU{2}_L$ and $\SU{3}_c$, respectively, and $\mu$ is the scale where the experiment actually measures the values of the gauge couplings, i.e.~in most cases $\mu=M_Z$. The $b_i$ are the $\beta$-function coefficients to be introduced below.

The $\beta$-function for a general gauge theory is given by the famous formula \cite{Politzer:1973fx,Gross:1973id}
\begin{align}
\beta(g) = -\frac{1}{16\pi^2} \big[ \frac{11}{3} \ell(&\text{vector}) - \frac{2}{3} \ell(\text{Weyl fermion})\notag\\ 
  &-  \frac{1}{6} \ell(\text{spinless}) \big] g^3 + \ldots,
\label{eq:beta-function}
\end{align}
where the dots indicate higher order corrections. In supersymmetric theories, this expression simplifies\footnote{There is a gaugino for each vector boson and 2 real bosons for each Weyl fermion, so combining the first and last 2 terms in Eq.~(\ref{eq:beta-function}), respectively, gives the 2 terms in Eq.~(\ref{eq:beta-function-susy}).} to
\begin{equation}
\beta(g) = -\frac{1}{16\pi^2} \overbrace{\left[3 \ell(\text{vector}) - \ell(\text{chiral}) \right]}^{\displaystyle b_i} g^3 + \ldots,
\label{eq:beta-function-susy}
\end{equation}
where ``vector'' and ``chiral'' denote the respective supermultiplets and $\ell(\ldots)$ is the index of the representation \cite{Slansky:1981yr}\footnote{In Ref.~\cite{Slansky:1981yr}, the definition of the index has to be amended by a factor of 1/2 in order to be consistent with Eq.~(\ref{eq:beta-function}) and most other definitions in the mathematics literature.},
\begin{equation}
\ell(\Lambda) = \frac{1}{2}\frac{\text{dim}(\Lambda)}{\text{dim}(\mathfrak{g})} \langle \Lambda, \Lambda+2\delta \rangle
\label{eq:dynkin-index}
\end{equation}
given in terms of the gauge group $\mathfrak{g}$, the highest weight $\Lambda$ of the representation in the Dynkin basis, and $\delta = (1,\ldots,1)$. The angular brackets $\langle \cdot, \cdot\rangle$ denote the scalar product.

\subsection{The New Particles}

For the MSSM particle content (including the gauge bosons which we have not listed in Tab.~\ref{tab:smparticles}), Eqs.~(\ref{eq:beta-function-susy}-\ref{eq:dynkin-index}) give
\begin{equation}
b_1 = -\frac{33}{5}, \qquad b_2 = -1, \qquad b_3 = 3,
\label{eq:beta-function-coefficients-mssm}
\end{equation}
from which e.g.~gauge coupling unification (see Fig.~\ref{fig:unif}) follows. 

\begin{table}
\renewcommand{\arraystretch}{1.7}
\caption{The {\it exotica} with their respective gauge groups. }
\label{tab:3-sets-of-exotics}
\begin{center}
\subtable[$\SU{3}_c\times\SU{2}_L\times\U{1}_Y$\label{tab:1st-set-of-exotics}]{\scalebox{0.85}{
\begin{tabular}{|c|l||c|l||c|l||c|l|}
\hline
$\check{Q}$ & $(\bs{3}, \bs{1})_{\q 1/3}$       &   $\check{E}_-$ & $(\bs{1}, \bs{1})_{\m 1}$   &  $\check{L}$ & $(\bs{1}, \bs{2})_{\q 0}$     &  $\check{E}_{\pm}$ & $(\bs{1}, \bs{1})_{\pm 1}$     \\
\hline
$\check{Q}^c$ & $(\bsb{3}, \bs{1})_{\m 1/3}$       &   $\check{E}_-^c$ & $(\bs{1}, \bs{1})_{\q 1}$   &  $\check{L}^c$ & $(\bs{1}, \bs{2})_{\q 0}$     &  $\check{E}_{\pm}^c$ & $(\bs{1}, \bs{1})_{\mp 1}$     \\
\hline
\end{tabular}}}

\subtable[$\text{SU}(3)_c \times \text{SU}(2)_L \times \text{SU}(2)_{A} \times \text{U}(1)_Y$]{\scalebox{0.84}{
\begin{tabular}{|l|l|l|l|}
\hline
$2\times(\bsb{3}, \bs{1}, \bs{1})_{\m 1/3}$ & $2\times(\bs{1}, \bs{1}, \bs{1})_{\q 1}$ & $2\times(\bs{1}, \bs{2}, \bs{1})_0$       & $2\times(\bs{1}, \bs{1}, \bs{1})_{\pm 1}$ \\
\hline
$2\times(\bs{3}, \bs{1}, \bs{1})_{\q 1/3}$   & $2\times(\bs{1}, \bs{1}, \bs{1})_{\m 1}$ & $2\times(\bs{1}, \bs{2}, \bs{1})_0$     & $2\times(\bs{1}, \bs{1}, \bs{2})_{\pm 1}$\\
\hline
\end{tabular}}}

\subtable[$\text{SU}(3)_c \times \text{SU}(2)_L \times \text{SU}(2)_{A} \times \text{SU}(2)_{B} \times \text{U}(1)_Y$]{\scalebox{0.75}{
\begin{tabular}{|l|l|l|l|}
\hline
$(\bs{3}, \bs{2}, \bs{1}, \bs{1})_{\q 1/3}$ & $5 \times (\bsb{3}, \bs{1}, \bs{1}, \bs{1})_{\q 2/3}$ & $2 \times (\bs{1}, \bs{2}, \bs{2}, \bs{1})_{\q 0}$ & $4 \times (\bs{1}, \bs{1}, \bs{2}, \bs{1})_{\pm 1}$ \\
\hline
$(\bsb{3}, \bs{2}, \bs{1}, \bs{1})_{\m 1/3}$   & $5 \times (\bs{3}, \bs{1}, \bs{1}, \bs{1})_{\m 2/3}$  &  $2 \times (\bs{1}, \bs{2}, \bs{1}, \bs{2})_{\q 0}$ & $4 \times (\bs{1}, \bs{1}, \bs{1}, \bs{2})_{\pm1}$ \\
\hline
\end{tabular}}}
\end{center}
\end{table}

Consider now the exotic particles given in Tab.~\ref{tab:1st-set-of-exotics}. We will denote their contribution to the $b_i$ by $\Delta b_i$, and, as an example, calculate $\Delta b_3$. In this case, the gauge group $\mathfrak{g}$ is $\SU{3}$ and the only representations which contribute are $\check{Q}$ and $\check{Q}^c$ in Tab.~\ref{tab:1st-set-of-exotics}. The highest weight of e.g.~$\check{Q}$ is $\Lambda = (1\,\, 0)$, and keeping in mind that the Dynkin scalar product is given by the quadratic form which is the inverse of the Cartan matrix \cite{DiFrancesco:1997nk}, Eq.~(\ref{eq:dynkin-index}) yields
\begin{equation}
\Lambda(\check{Q}) = \frac{1}{2} \cdot \frac{3}{8} \cdot (1\,\,0) \frac{1}{3}\begin{pmatrix}2 & 1\\ 1 & 2 \end{pmatrix} \begin{pmatrix} 3\\ 2 \end{pmatrix} = \frac{1}{2}.
\label{eq:calculate-b3}
\end{equation}
$\check{Q}^c$ contributes the same amount, so in the end we have $\Delta b_3 = -1$, see Eq.~(\ref{eq:beta-function-susy}). 

The interesting point now is that for this special set of exotics, all $\Delta b_i$ are equal:
\begin{equation}
\Delta b_3 = -1, \quad \Delta b_2 = -1, \quad \Delta b_1 = -1
\label{eq:bi-1st-set-exotics}
\end{equation}
At the same time, the states in Tab.~\ref{tab:1st-set-of-exotics} are {\it not} in a $\bs{5}$ or $\bs{10}$ of $\SU{5}$. In the following section \ref{sec:predictions-from-guts}, we will show that most predictions from Grand Unification are {\it not changed} in the presence of these states.

\subsection{The Predictions from Grand Unification}
\label{sec:predictions-from-guts}

Eq.~(\ref{eq:rge}) really corresponds to 3 equations (one for each value of the index $i=1,2,3$) with 1 unknown $\mgut$. We can e.g.~take the first 2 equations, solve for $\mgut$, use the relations $\alphaem = \alpha_2 \sin^2\theta = 3/5\, \alpha_1 \cos^2\theta$ and substitute $\mu=M_Z$:

\vspace{-3ex}
\subsubsection*{\ding{202} The \gut{} Scale}
\begin{equation}
\textstyle
\mgut = M_Z \, \exp\left[\frac{2\pi}{\Red{b_2 - b_1}} \frac{1}{\alphaem(M_Z)}\left(\frac{3}{5} - \frac{8}{5} \sin^2\theta(M_Z) \right) \right]
\end{equation}
It turns out that $\mgut$ only depends on the {\it difference} between the $b_i$. Since our exotica contribute equally to the $b_i$ (see Eq.~(\ref{eq:bi-1st-set-exotics})), the value of $\mgut$ is {\it unchanged} in the presence of these states.

\vspace{-3ex}
\subsubsection*{\ding{203} The Strong Coupling Constant}

\begin{equation}
\textstyle
\frac{1}{\alpha_3(M_Z)} =  \frac{\sin^2\theta(M_Z)}{\alphaem(M_Z)} - \frac{\Red{b_3-b_2}}{2\pi}\log\left(\frac{\mgut}{M_Z} \right)
\end{equation}
Since $\alpha_3$ is less well measured than the other coupling constants $\alpha_2$ and $\alpha_1$, one usually solves for $\alpha_3$ at the lower scale in terms of $\mgut$ and compares the low-energy prediction to experiment. Again, the result depends only on the difference of the $b_i$ and thus cancels for the states in Tab.~\ref{tab:1st-set-of-exotics}.

\vspace{-3ex}
\subsubsection*{\ding{204} The Weinberg Angle}

At the \gut{} scale, $\sin^2\theta$ is determined by the relative normalization of $\U{1}_Y$ to the Cartan generator in the \gut{} group which corresponds to hypercharge, see e.g.~section 4.4 of Ref.~\cite{Raby:2007yc} for a detailed discussion using the notation of the present publication. The low energy prediction for $\sin^2\theta$ is not independent of the one for $\alpha_3$.

\vspace{-3ex}
\subsubsection*{\ding{205} The $\gut$ Coupling Constant}
\begin{equation}
\textstyle 
\frac{1}{\alphagut} = \frac{\sin^2\theta(M_Z)}{\alphaem(M_Z)} + \Red{\frac{b_2}{b_2-b_1}} \frac{1}{\alphaem(M_Z)}\left(\frac{3}{5} - \frac{8}{5}\sin^2\theta(M_Z) \right)
\label{eq:gut-coupling-constant}
\end{equation}
The \gut{} coupling constant is the only prediction from Grand Unification which does change (namely it increases), since it does not only depend on the difference $b_2-b_1$, but also directly on $b_2$.

\subsection{Connection to String Theory}

We found the 3 sets of exotica given in Tab.~\ref{tab:3-sets-of-exotics} in the 5th twisted sector of the $\mathbb{Z}_6$ orbifold \cite{Kobayashi:2004ud,Kobayashi:2004ya,Buchmuller:2005jr}. As a matter of fact, the states in Tab.~\ref{tab:1st-set-of-exotics} transform as
\begin{equation}
(\bs{6},\bs{1}) + (\bs{1},\bs{2}) + \text{ c.c}
\end{equation}
under $\SU{6}\times\SU{2}$, before this symmetry is broken by Wilson lines to Pati-Salam and then spontaneously to the Standard Model.

In the minilandscape search \cite{Lebedev:2006kn,Lebedev:2007hv}, we derived 127 MSSM-like models from string theory. It turns out that $\sim$ 5\% of these models contain exotics of the type listed in Tab.~\ref{tab:3-sets-of-exotics}.

\section{Phenomenological Consequences}

\subsection{Exotic Mesons and Baryons}

Consider the exotica in Tab.~\ref{tab:1st-set-of-exotics} and notice that their electric charge $Q=T_3+Y/2$ is fractional. As a consequence, the bound states of exotic quarks and SM quarks (see Tab.~\ref{tab:exotic-mesons-baryons}) will also have fractional electric charge.

We will assume that the exotic quarks have a gauge invariant, supersymmetric mass $M$ of order the electroweak scale. Searches for fractionally charged heavy particles exclude them with mass less than 200 GeV \cite{Acosta:2002ju,Perl:2004qc,Fairbairn:2006gg}. Nevertheless they can be produced at the Tevatron or the LHC via Drell-Yan processes. For more details, cf.~Ref.~\cite{Raby:2007hm}.

\begin{table}
\renewcommand{\arraystretch}{1.7}
\caption{The exotica form (fractionally charged) mesons and baryons. The superscripts denote electric charge. }
\label{tab:exotic-mesons-baryons}
\begin{center}
\subtable[Exotic mesons.]{\scalebox{0.8}{
\begin{tabular}{|l||l|}
\hline
$Q_u^{+1/2}$         & $Q_d^{-1/2}$\\
\hline
$[\check{\bar Q} u]_{+1/2}$  & $[\check{\bar Q} d]_{-1/2}$\\
\hline
\end{tabular}}}

\subtable[Exotic baryons.]{\scalebox{0.8}{
\begin{tabular}{|l||l||l||l|}
\hline
$\Sigma_Q^{+3/2}$ & $\Sigma_Q^{+1/2}$   & $\Sigma_Q^{-1/2}$  & $\Lambda_Q^{+1/2}$\\
\hline
$[\check{Q} u u]_{3/2}$   & $[\check{Q} (u d)_s]_{1/2}$ & $[\check{Q} d d]_{-1/2}$   & $[\check{Q} (u d)_a]_{1/2}$\\
\hline
\end{tabular}}}

\end{center}
\end{table}

\subsection{Exotica with Hidden Sector Charge}

In string theory the exotica may transform non-trivially under a hidden sector
gauge group. Here we consider a generalized example, not obtained
from a particular string construction, which has interesting
phenomenology. Consider a hidden sector gauge group $\text{SU}(N)$
with the exotica transforming as
\begin{equation} [(\bs{6}, \bs{1}, \bs{N}) + (\bs{1}, \bs{2}, \bs{N})] + \text{c.c.}  \label{eq:Nexotica2} \end{equation}
under $\text{SU}(6) \times \text{SU}(2)_R \times \text{SU}(N)$.
Note that values of $N > 3$ generically give too large a value for $\alphagut$ (see Eq.~(\ref{eq:gut-coupling-constant}) and the following discussion) and are thus excluded by demanding perturbative unification.

Assuming the hidden sector gauge coupling gets strong at a
scale $\Lambda_N \gg M_Z$, the exotica will form $\text{SU}(N)$ singlet
bound states with mass of order $\Lambda_N$, and the
phenomenology of such $\text{SU}(N)$ singlet ``baryons" and ``mesons" will
depend on the values of $N$ and $\Lambda_N$.

Current bounds restrict us to a gauge invariant mass $M \gtrsim 200$. We then can consider two possibilities, either $\Lambda_N \geq M$ or $\Lambda_N \ll M$.   The first case is comparable to
QCD with all quark masses less than or equal to
$\Lambda_{\text{QCD}}$. The second case is more interesting.
The exotica will have properties
similar to the ``quirks" introduced in Ref.~\cite{Kang:2006yd}. They
can be produced at the LHC.  When the exotics are produced they can
separate by large distances in the detector before forming the bound
state, since their effective string tension is so much smaller than
their mass. Again, cf.~Ref.~\cite{Raby:2007hm} for more details.

\section{Conclusions}

Supersymmetry and Grand Unification are prime candidates for physics beyond the SM. As such, we would very much like to keep their predictions as guidelines for model building. It is well known that states that come in complete multiplets of the grand unified gauge group do not affect gauge coupling unification. In this study, we have discussed a novel feature of light exotic particles
\begin{enumerate}[(i)]
\item that do not come in complete $\text{SU}(5)$ multiplets,
\item that do not affect gauge coupling unification and most other predictions of \gut{}s (at one loop),
\item whose only effect is to increase the value of $\alphagut$,
\item that have fractional electric charge,
\item that are found in orbifold constructions of the heterotic string.
\end{enumerate}
Clearly, all these exotica would have very distinctive signatures at the LHC.

\textbf{Acknowledgments.} We are indebted to the Ohio Supercomputing Center for using their resources and to Ben Dundee and Hasan Y\"uksel for useful discussions. We would like to acknowledge research supported in part by the Department of Energy under Grant No.~DOE/ER/01545-872.


\end{document}